# High-Pressure XRD Study of Ti-3Al-2.5V Titanium Alloy: Intermediate Transition Pressure and Composition Trends in Ti–Al–V Alloys

D. Errandonea[1], R. Turnbull[1], P. Botella[1], R. Oliva[2], C. Popescu[3], S. MacLeod[4]

[1]Department of Applied Physics - ICMUV, Universitat de València, Av. Vicent Andres Estelles 19, 46100 Valencia, Spain

[2]Geosciences Barcelona (GEO3BCN-CSIC), C/ Lluís Solé i Sabarís s/n, 08028 Barcelona, Catalonia, Spain

[3]CELLS-ALBA Synchrotron Light Facility, Cerdanyola del Vallès, Barcelona 08290, Spain

[4]AWE, Aldermaston, Reading, RG7 4PR, United Kingdom

**Abstract:** High-pressure X-ray diffraction experiments were performed on Ti–Al–V alloys to investigate the effect of composition on structural stability, focusing on Ti-3Al-2.5V and comparing with pure Titanium and Ti-6Al-4V. Measurements using different pressure-transmitting media show a phase transition in Ti-3Al-2.5V at 17–19 GPa, intermediate between pure Ti (5–10 GPa) and Ti-6Al-4V (~30 GPa). Despite variations arising from the choice of pressure medium, the transition pressure shows a clear and systematic increase with higher Al and V content. Equation-of-state analysis indicates that the bulk modulus remains nearly unchanged across compositions. This suggests a decoupling between elastic properties and phase stability, with alloying primarily affecting the transition pressure rather than compressibility. These results highlight the role of composition in tuning high-pressure phase transformations in Ti-Al-V based alloys.

## Introduction

Titanium/Aluminum/Vanadium (Ti-Al-V) alloys were hailed as a breakthrough with strategic industrial significance. Alloys such as Ti-6Al-4V (Ti64) and Ti-3Al-2.5V (Ti3-2.5) are recognized as critical materials for aerospace and biomedical applications due to their exceptional strength-to-weight ratio and high-temperature performance. They are the most commercially successful titanium alloys, having shaped numerous industrial and commercial applications, ranging from jet engine components to medical implants [1]. Ti64 and Ti3-2.5 are also used in components of Fusion Reactors [2]. This is because they exhibit fast decay rates for residual radioactivity, which is crucial for enabling hands-on maintenance of components after the reactor is shut down and reducing long-term waste. These alloys also show good compatibility with coolants such as lithium, helium, and water. Ti-Al-V alloys such as Ti64 and Ti3-2.5 are considered for fusion applications due to their microstructural and mechanical stability under external stress. Their deformation mechanisms and fatigue behavior, which are closely linked to phase constitution and mechanical properties, have been extensively studied in the context of fusion reactor materials [3]. Moreover, the dependence of fatigue performance on microstructure and defects highlights the importance of elastic and mechanical properties, including modulus-related parameters, in preventing failure [4].

High-pressure (HP) studies using a diamond-anvil cell (DAC) are among the best suited methods to characterize the structural stability and mechanical properties of metals and alloys [5, 6, 7]. Previous HP studies have been performed on Titanium (Ti) [8, 9] and Ti64 [10, 11] combining X-ray diffraction (XRD) with the use of a DAC. Those experiments were carried out using a variety of pressure-transmitting media (PTM). A summary of the results is schematically shown in Figure 1. In the former studies, it was found that the incorporation of Al and V in Ti64 increases structural stability. The transition pressure from $\alpha$-to-$\omega$ was found to increase from near 10 GPa in Ti to around 30 GPa in Ti64. The transition pressure and the bulk modulus might be dependent not only on the inclusion of Al and V, but also on the experimental conditions. For instance, in the case of Ti, a seminal work using different PTM demonstrated that non-hydrostatic stresses lower the transition pressure from 11 to 5 GPa [8]. They also were found to affect the reversibility of the phase transition [8]. A systematic comparison of other Ti-Al-V alloys, like Ti3-2.5, with Ti64 and pure Ti is needed to better understand how the concentration of Al and V affects the properties of the Ti-Al-V alloys under extreme conditions. Unfortunately, HP studies on Ti3-2.5 have never been performed. The aim of this work is to investigate the room-temperature structural evolution upon compression in Ti3-2.5 by combining XRD with the use of a DAC, using different PTM, to explore its compressibility, stability, and phase transformations up to a pressure of 30 GPa.

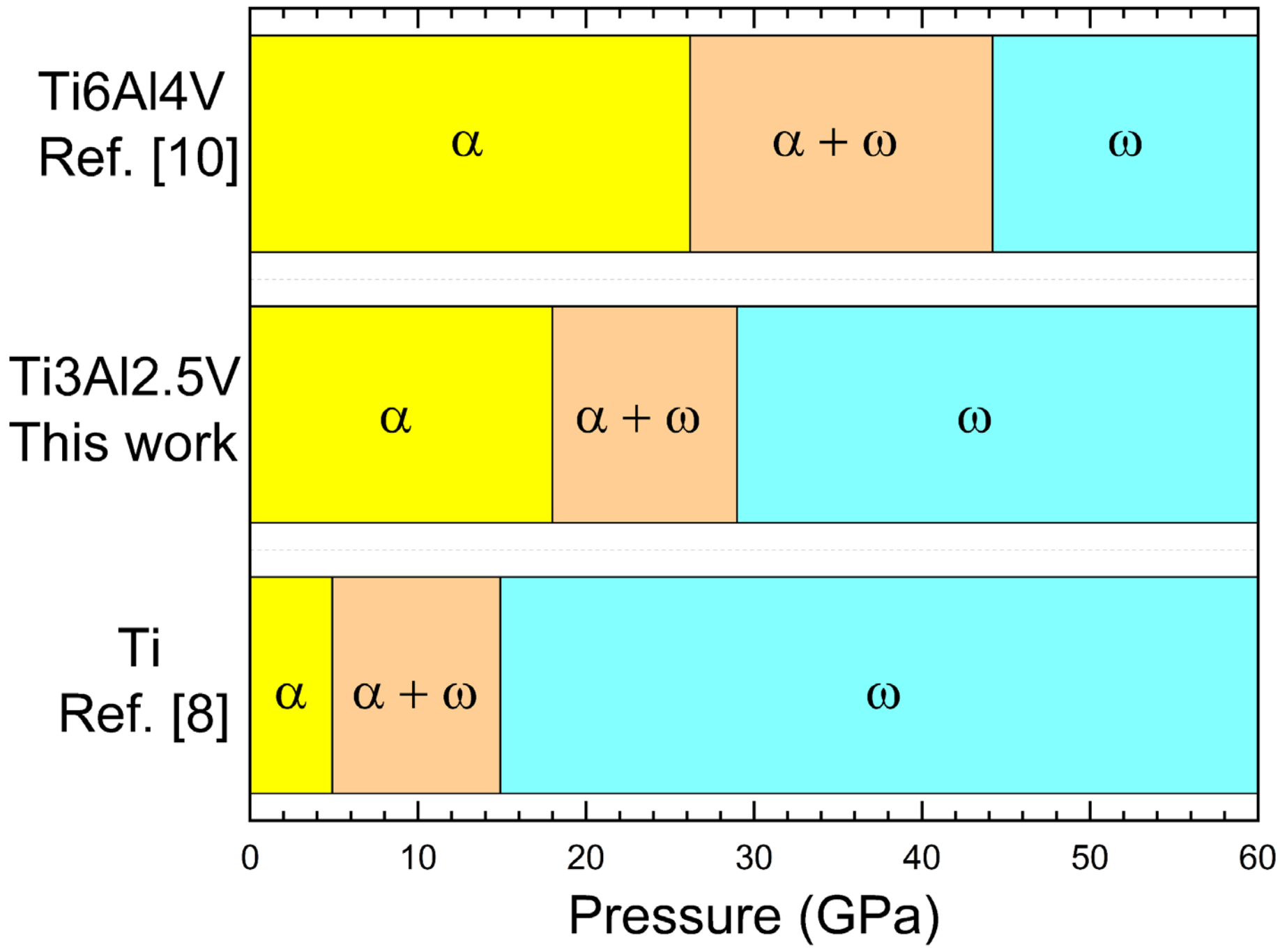

**Figure 1:** Schematic representation of the high-pressure room-temperature structural sequence of Ti and Ti-A-V alloys. For Ti, the figure compiles results obtained using argon, 4:1 methanol–ethanol, and NaCl as pressure-transmitting media, as well as experiments conducted without a pressure medium [8]. For Ti3Al2.5V, the figure summarizes the present results obtained using 4:1 methanol–ethanol, silicone oil, and MgO as pressure-transmitting media. For Ti6Al4V, the figure compiles results obtained using neon, 4:1 methanol–ethanol, and silicone oil as pressure-transmitting media, together with experiments performed without a pressure medium [10].

## Materials and methods

The samples employed for this study were obtained from a polycrystalline high-purity Ti3Al2.5V foil (5 μm thick) purchased from Goodfellow Advanced Materials. The crystal structure was confirmed by XRD to be stable in the α-phase (hcp) [8, 10], space group $P6_3/mmc$, and with unit-cell parameters $a$ = 2.932(1) Å and $c$ = 4.672(2) Å. Figure 2 presents the Rietveld refinement of XRD data collected at ambient conditions using a Malvern PANalytical X'Pert Pro diffractometer with Cu $K\alpha_1$ radiation (λ = 1.5406 Å). Modeling the crystal structure in the α-phase yielded a good fit to all observed peaks, with goodness-of-fit parameters $R_p$ = 4.16%, $R_{wp}$ = 5.28%, and $\chi^2$ = 1.54.

We performed three independent high-pressure experiments using synchrotron XRD and a membrane DAC. In each experiment, a fresh sample and a different PTM was used: a 4:1 methanol–ethanol (ME) mixture, silicone oil (SO), and magnesium oxide (MgO). We performed one experimental run for each PTM. These three media are all commonly employed in DAC studies, but they exhibit markedly different behavior under pressure, which influences the degree of quasi-hydrostaticity of the experiments. The ME mixture remains hydrostatic up to ~10 GPa, above which it solidifies and generates non-hydrostatic stresses [12]. SO undergoes a similar solidification at a lower pressure of ~4 GPa and develops more pronounced non-hydrostatic conditions than ME [12]. In contrast, MgO is solid at ambient conditions and not fully hydrostatic at low pressures, but at pressures above ~20 GPa - where all liquid PTMs would have already become solids - it provides more reasonable quasi-hydrostatic conditions [13].

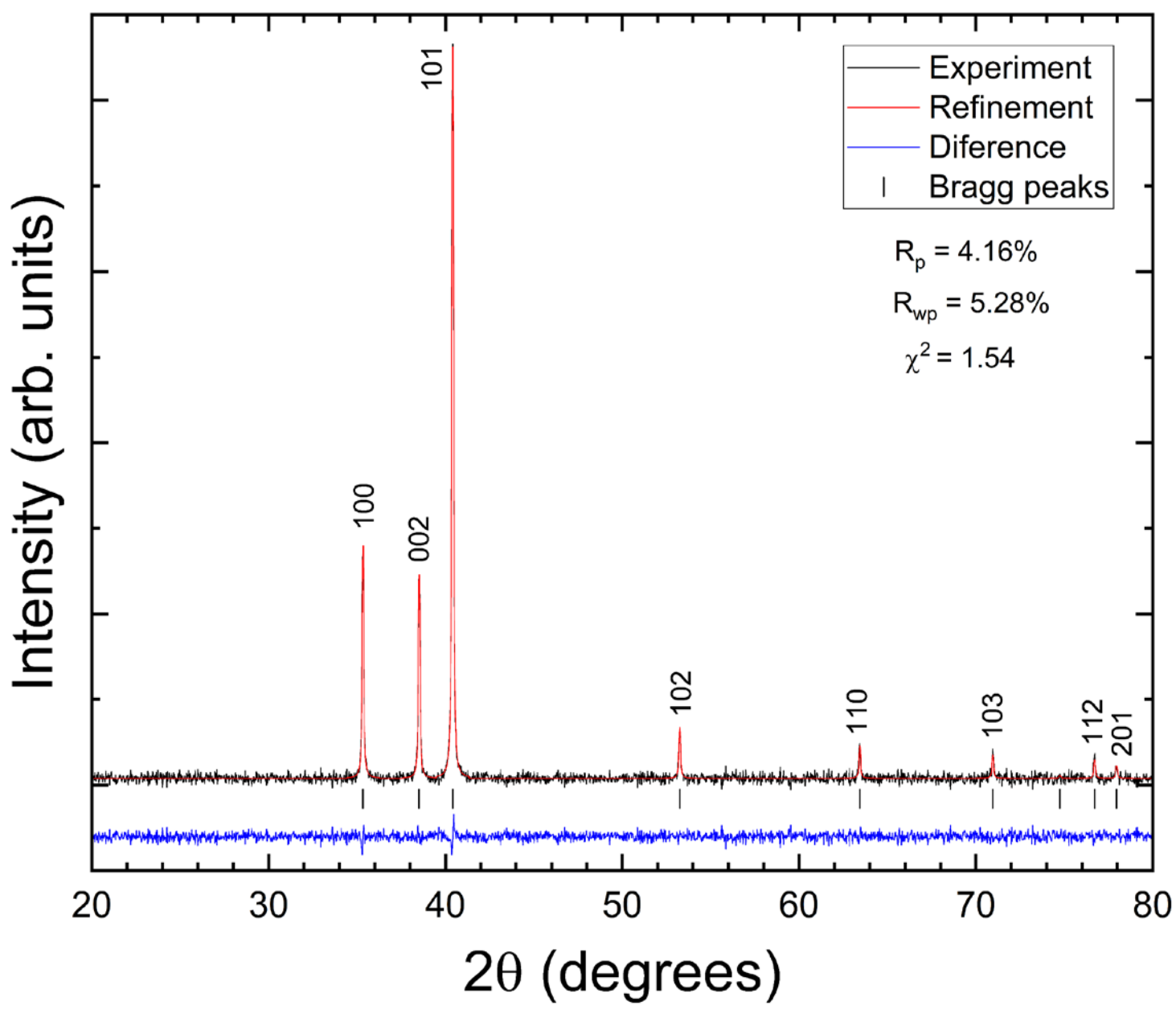

**Figure 2**: XRD pattern collected from Ti3Al2.5V at ambient conditions (λ = 1.5406 Å). The figure includes the Rietveld refinement and goodness-of-fit parameters.

High-pressure angle-dispersive powder XRD measurements were performed at room temperature at the BL04-MSPD beamline of the ALBA synchrotron facility [14]. Experiments were conducted using a wavelength of 0.4246 Å and a Rayonix SX165 CCD detector. The sample, along with a copper grain as a pressure standard [15], was loaded into three separate DACs equipped with 400 μm-diameter diamond culets, each containing one of the pressure-transmitting media described above. During loadings in the DACs, the Cu grains were positioned approximately 3 μm away from the sample, toward the gasket. At each pressure we collected one pattern with sample and Cu (used to measure pressure) and another with only the sample (used for structure identification). The sample-to-detector distance was calibrated by fitting a $LaB_6$ standard using the Dioptas software, which was also employed to process the 2D diffraction images into conventional 1D diffractograms. Structural identification and lattice parameter determination as a function of pressure were performed using the Le Bail method [16] implemented in the FullProf package.

## Results and discussion

The waterfall plot in Figure 3 shows a selection of integrated XRD patterns of Ti3-2.5 embedded with ME acting as the PTM. It includes XRD patterns collected under compression, along with one pattern recorded after decompression. The patterns were analyzed using the Le Bail method to account for the presence of preferred orientations, which did not prevent the identification of the $\alpha$ and $\omega$ phases or the determination of their unit-cell parameters. Due to the limited angular aperture of the DAC used in the experiment carried out under ME, the accessible range was restricted to $2\theta < 16°$ (compared to $2\theta < 25°$ in the other two experiments). However, within this range, four diffraction peaks corresponding to the $\alpha$ and $\omega$ phases were observed, which are sufficient for phase identification and reliable determination of the lattice parameters.

At ambient pressure and pressures below 18.60(5) GPa all the XRD patterns we collected are in the $\alpha$-phase, space group $P6_3/mmc$. At this pressure, a shoulder was observed around 11°, located between the $\alpha$-phase peaks with Miller indices 002 and 101. This additional feature is indicated by an arrow in Figure 3. The inset of the figure shows an expanded view of the XRD pattern between 10.5° and 12°, highlighting the emergence of the extra peak, which corresponds to the strongest reflections, the (110/101), of the $\omega$-phase, space group $P6/mmm$. The change in the XRD pattern at 18.60(5) GPa serves as a fingerprint of the $\alpha \rightarrow \omega$ transition. As pressure increases, the peaks of the $\omega$-phase grow in intensity while those relating to the $\alpha$-phase decrease. This can be seen in the XRD pattern at 22.70(5) GPa. The transition is completed at 29.8(1) GPa. At this pressure only the $\omega$-phase peaks are present. We also performed a decompression to assess the reversibility of the $\alpha \rightarrow \omega$ phase transition. The lowest pressure reached was 3.10(2) GPa, as friction between the piston and cylinder of the DAC prevented further pressure release. At this pressure, the sample had not reverted to the $\alpha$-phase, which appears to be consistent with the findings from experiments carried out on Ti using the same PTM [8]. It is important to comment here that although the coexistence of α and ω phases indicates a progressive transformation, the diffraction patterns were affected by preferred orientation effects (in the three experiments), which are common in DAC studies on metallic samples undergoing martensitic transformations. Therefore, the relative peak intensities cannot be reliably used to quantify phase fractions or determine the detailed evolution of the transformation as a function of pressure. The quantitative determination of phase fractions would require experiments specifically optimized to minimize texture effects.

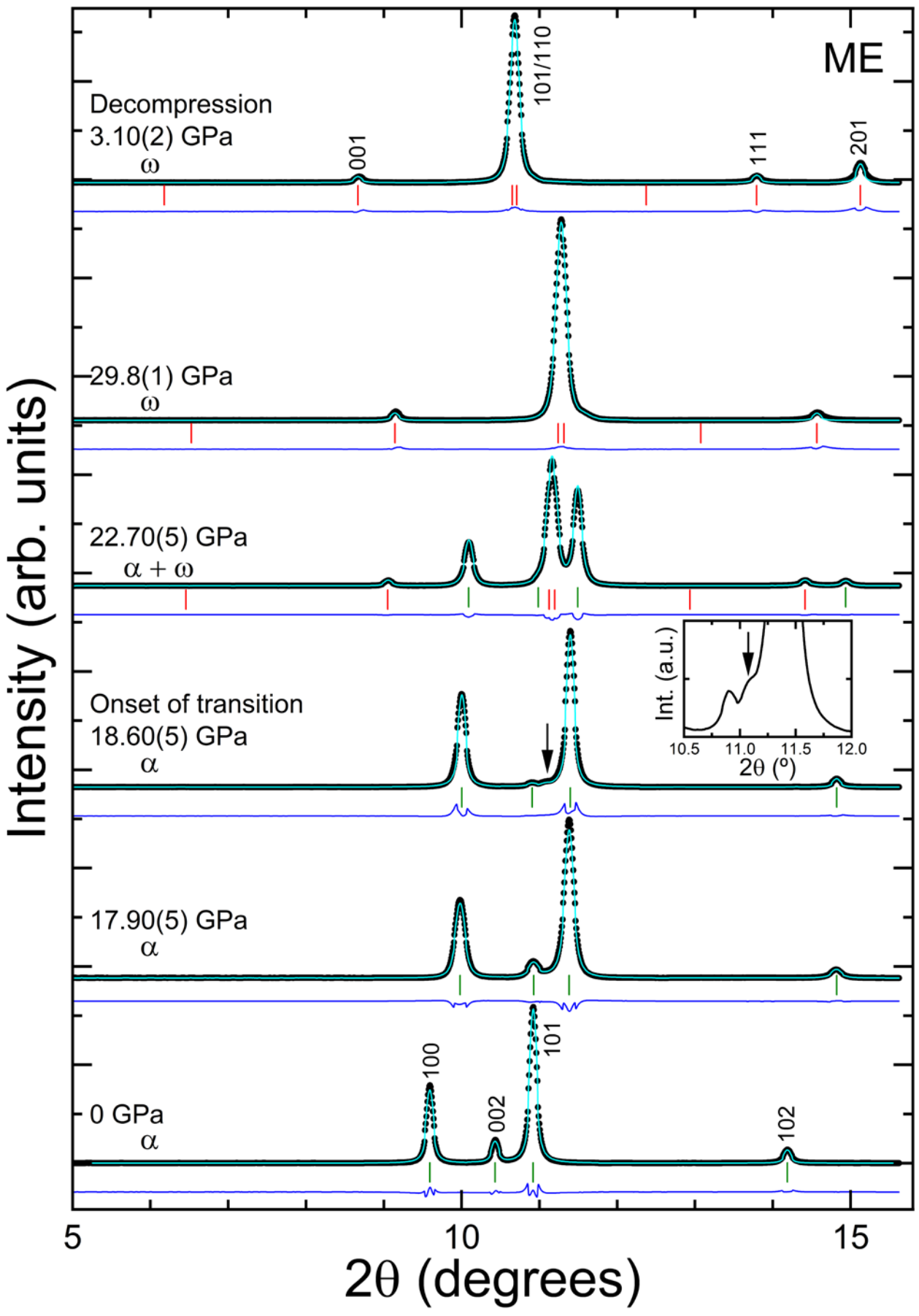

**Figure 3:** XRD patterns of Ti3-2.5 embedded in an ME PTM, $\lambda$ = 0.4246 Å. Black circles represent experimental data. Cyan color lines are the Le Bail fits. Blue lines are the residuals of the fits. Vertical green (red) ticks show the calculated positions for peaks of the α-phase (ω-phase). Miller indices of the peaks of the α-phase (ω-phase) are included in the bottom (top) trace of the figure. The arrows depict the emergence of the first peak of the ω-phase. The inset shows the 10.5º-12.0º region of the XRD pattern measured at 18.60(5) GPa to highlight the emergence of the strongest peak of the ω-phase.

Figure 4 shows results from the experiment performed using silicone oil (SO) as the PTM. In this experiment the onset of the $\alpha \rightarrow \omega$ transition in Ti3-2.5 is detected at 17.10(5) GPa. The emergence of the ω-phase is not only identified by the appearance of an extra peak between the 002 and 101 peaks of the $\alpha$-phase but also by two other peaks. The peaks of the $\omega$-phase at 17.10(5) GPa are identified by arrows in the figure. Upon further compression, changes in the XRD pattern are similar to the changes observed in the experiment performed under ME. The peaks of the $\omega$-phase increase in intensity whilst those of the $\alpha$-phase decrease under compression. A pure $\omega$-phase was found at 30.0(1) GPa. Upon decompression we found a partial reversion of the phase transition. The lowest pressure measured was 5.10(3) and both phases were present at this pressure. Thus, the $\alpha \rightarrow \omega$ transition is at least partially reversible in SO. This finding is consistent with results from experiments carried out on Ti where it was also found that the PTM could affect the reversibility of the transition [8]. In Ti, in the more quasi-hydrostatic experiments (argon or ME as PTM), the transition was irreversible and in the less quasi-hydrostatic experiments (NaCl as PTM or no PTM) the transition was partially reversible.

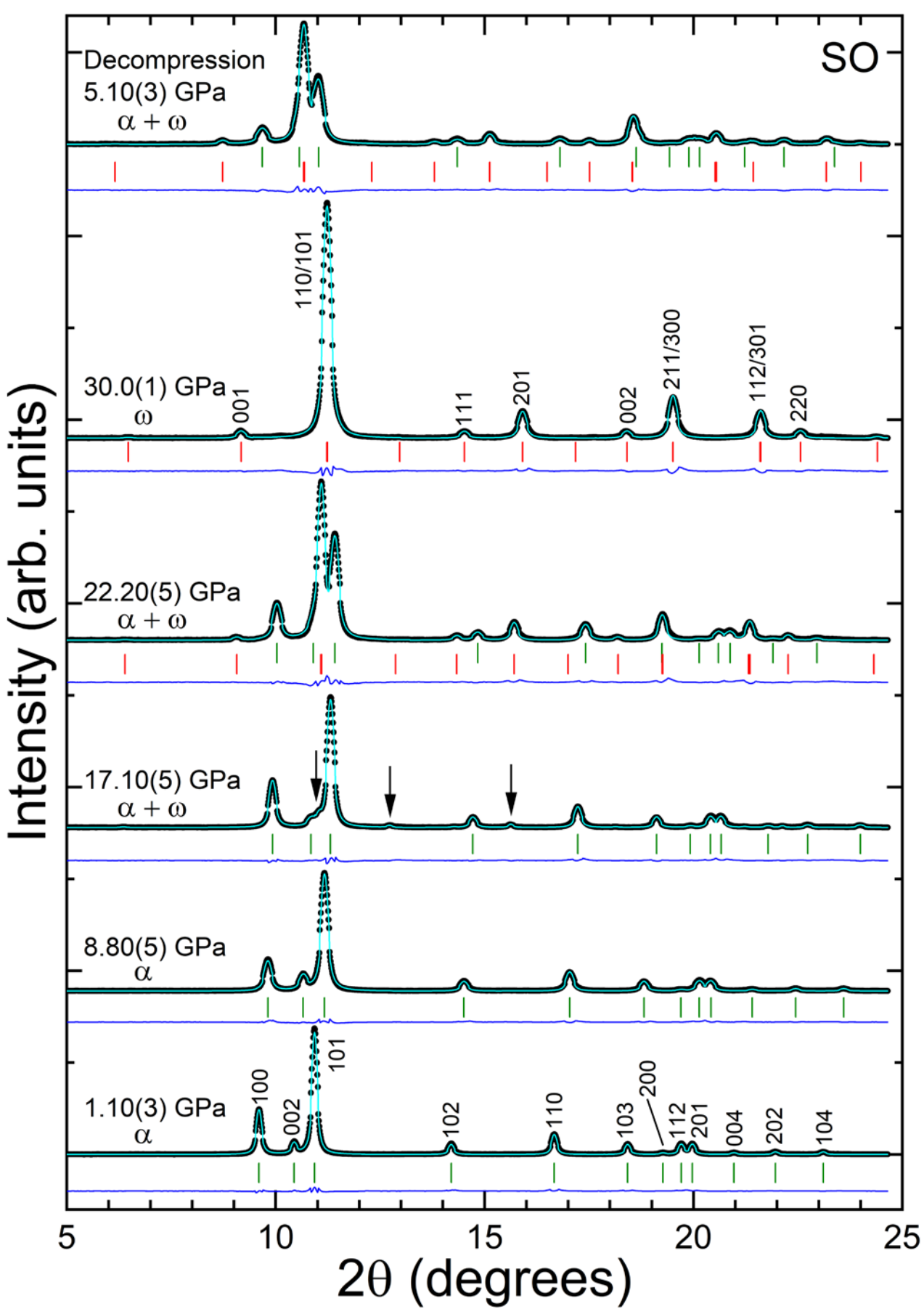

**Figure 4:** XRD patterns of Ti3-2.5 embedded in a SO PTM, $\lambda$ = 0.4246 Å. Black circles represent experimental data. Cyan color lines are the Le Bail fits. Blue lines are the residuals of the fits. Vertical green (red) ticks show the calculated positions for peaks of the $\alpha$-phase ($\omega$-phase). Miller indices of the peaks of the $\alpha$-phase and $\omega$-phase are included in the XRD patterns measured at 1.10(3) GPa and 30.0(1) GPa, respectively. The arrows depict the emergence of the first peaks of the $\omega$-phase.

The results obtained using MgO as the PTM are qualitatively similar to those from the two previously described experiments (PTM = ME or SO). Figure 5 presents the results obtained using MgO as PTM. In this experiment, we also detected the peaks from MgO which do not prevent the identification of the $\alpha$ and $\omega$ phases. The peaks from MgO [13] were also used to confirm the pressure determined from the Cu standard. In the MgO experiment, the onset of the $\alpha \rightarrow \omega$ transition is detected at 18.90(5) GPa. As in the ME experiment, the emergence of the $\omega$-phase is also identified by the appearance of an additional peak between the 002 and 101 reflections of the $\alpha$-phase, which is identified by an arrow in the figure. The figure also includes an inset of the region of interest to facilitate the identification of the extra peak.

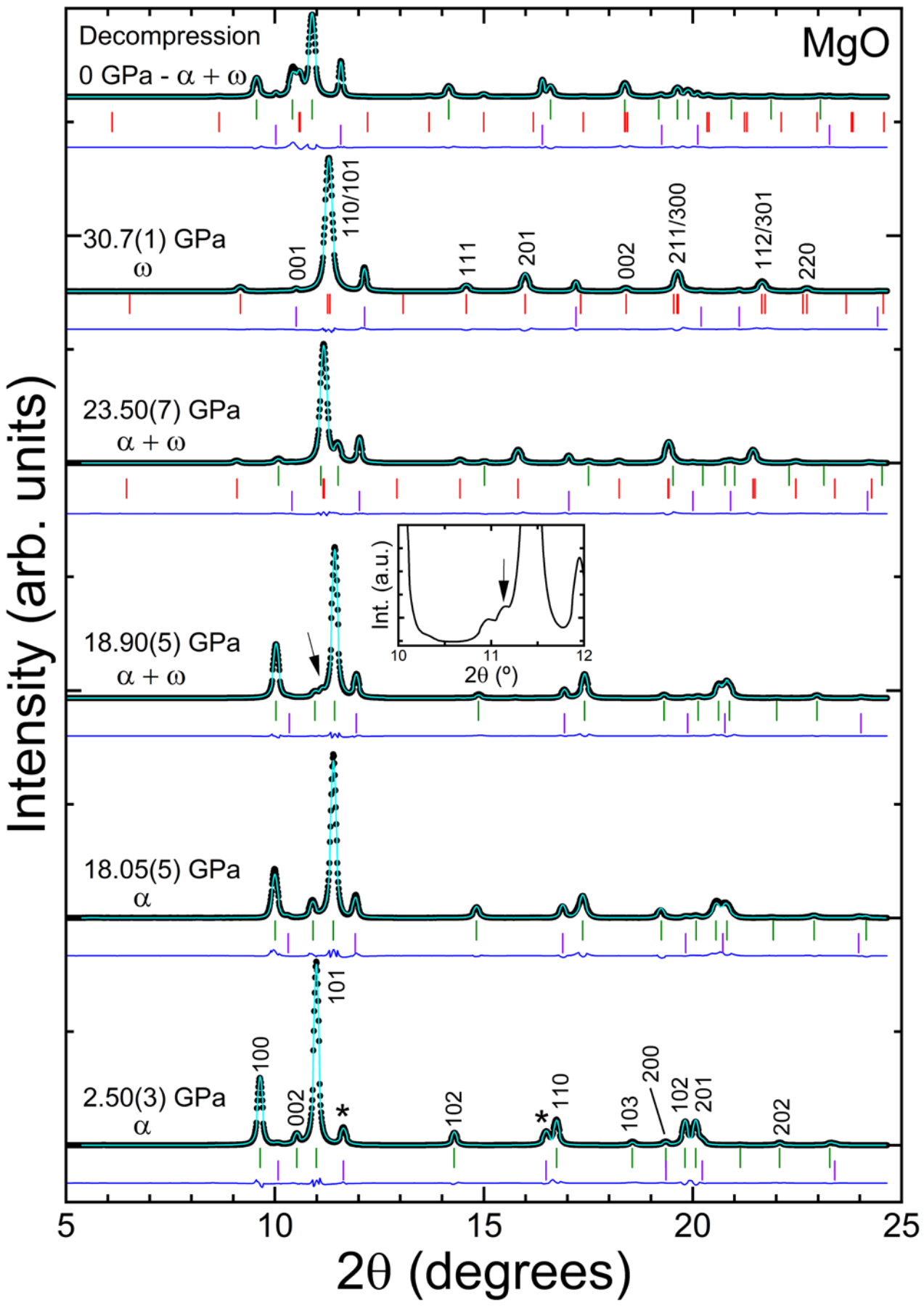

**Figure 5:** XRD patterns of Ti3-2.5 embedded in an MgO PTM, $\lambda$ = 0.4246 Å. Black circles represent experimental data. Cyan color lines are the Le Bail fits. Blue lines are the residuals of the fits. Vertical green, red, and brown ticks show the calculated positions for peaks of the $\alpha$-phase, $\omega$-phase, and MgO, respectively. Miller indices of the peaks of the $\alpha$-phase and $\omega$-phase are included in XRD patterns measured at 2.50(3) GPa and 30.7(1) GPa, respectively. The arrows depict the emergence of the first peak of the $\omega$-phase. The inset shows a zoom of the 10º-12º region of the XRD pattern measured at 18.90(5) GPa to facilitate the identification of the emergence of the strongest peak of the $\omega$-phase.

When pressure increases further, the evolution of the XRD patterns is similar to that observed in the ME and SO experiments. The intensity of the peaks of the ω-phase is enhanced, while the peaks from the α-phase progressively decrease in intensity. A fully transformed ω-phase is observed at 30.7(1) GPa. During decompression, we found a partial reversion of the phase transition. In this experiment we reached ambient pressure during decompression. At this pressure both α- and ω-phases coexist, as observed in the SO experiment.

In Table 1 and Figure 1, we summarize the results of this work and compare them with previous studies in Ti [8] and Ti64 [10]. The present results, combined with previous high-pressure studies on Ti and Ti64, reveal a clear and systematic evolution of structural stability across the Ti–Al–V alloy system. This trend is summarized in Figure 6, where the transition pressure is plotted as a function of alloy composition. The $\alpha \rightarrow \omega$ transition pressure increases progressively from pure Ti (~5–10 GPa) to Ti3Al2.5V (~17–19 GPa) and further to Ti64 (~26–32 GPa), demonstrating that alloying with Al and V stabilizes the α-phase against pressure-induced transformation. Overall, the data reveal a monotonic increase in transition pressure with increasing alloying content in the Ti–Al–V alloys investigated here. However, because the Al/V ratio is not independently varied among the studied compositions, the present results do not allow the individual contributions of Al and V, or the specific effect of their ratio, to be distinguished.

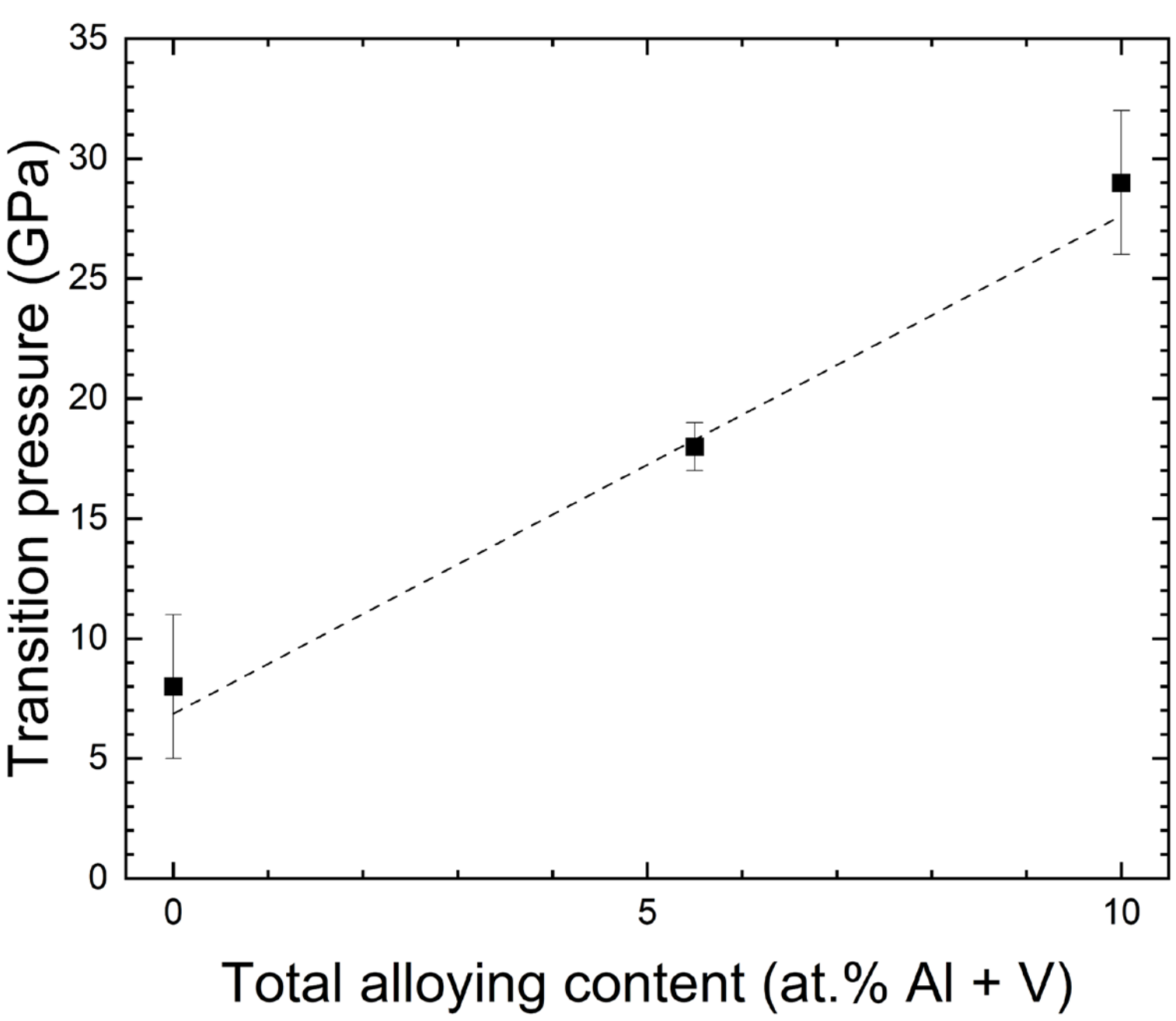

**Figure 6:** α → ω transition pressure as a function of alloy composition in Ti–Al–V alloys. Data for Ti and Ti-6Al-4V are taken from Refs. [8,10], while Ti-3Al-2.5V corresponds tothis work. The dashed line is a guide to the eye, highlighting the monotonic increase in transition pressure with alloying content.

This trend can be understood in terms of the role of alloying elements in modifying both the electronic structure and lattice energetics. Aluminum is known to act as a $\alpha$-phase-stabilizer, strengthening the hexagonal close-packed (hcp) structure by increasing directional bonding and reducing the driving force for transformation [17]. Vanadium, although typically considered a stabilizer of the $\beta$-phase (space group $Im\bar{3}m$) at ambient conditions [18], contributes to solid-solution strengthening and alters the electronic density of states, which can influence phase stability under compression [17]. The combined presence of Al and V therefore increases the energy barrier between the $\alpha$ and $\omega$ phases, shifting the transition to higher pressures.

| Material | Pressure medium | Onset of transition | Reference |
| --- | --- | --- | --- |
| Ti | No medium | 4.9 GPa | [8] |
| | NaCl | 6.2 GPa | |
| | ME | 10.2 GPa | |
| | Argon | 10.5 GPa | |
| Ti3Al2.5V | MgO | 18.9 GPa | This work |
| | SO | 17.1 GPa | |
| | ME | 18.6 GPa | |
| Ti64 | No medium | 32.1 GPa | [10] |
| | Mineral oil | 26.2 GPa | |
| | ME | 31.2 GPa | |
| | Neon | 32.7 GPa | |

**Table 1:** The observed $\alpha$-$\omega$ phase transition pressure in Ti, Ti3Al2.5V, and Ti64.

Importantly, Ti3-2.5 exhibits a transition pressure that lies approximately midway between that of Ti and Ti64, indicating that the effect of alloying is approximately compositional and monotonic rather than abrupt. This suggests that the $\alpha$-phase stability can be continuously tuned by adjusting alloy composition, providing a useful framework for designing Ti-Al-V based alloys with tailored high-pressure performance.

Another important aspect is the role of non-hydrostatic stress, as evidenced by the dependence of transition pressure and reversibility on the pressure-transmitting medium. While this effect is also observed in pure Ti, its persistence in Ti3-2.5 indicates that stress conditions can compete with compositional effects in determining the observed transition pathway. Nevertheless, the overall compositional trend remains robust across different experimental conditions. In addition, the fraction of retained ω-phase upon decompression may depend on the maximum pressure reached prior to unloading, owing to hysteresis, defect accumulation, and incomplete reversibility of the martensitic transformation. Since decompression in the present experiments was only performed after reaching the maximum pressure (~30 GPa), this possible path dependence was not systematically investigated here.

The pressure-dependent evolution of the unit-cell parameters in the α and ω phases of Ti3-2.5 was analyzed based on XRD measurements, with the corresponding results summarized in Figure 7. Unit-cell parameters were not systematically analyzed during decompression as decompression data are often affected by hysteresis, non-hydrostatic stress, and irreversible structural changes (e.g., defect formation or metastable phase retention). These effects can cause deviations from equilibrium behavior, such that the decompression path does not accurately reproduce the intrinsic pressure–volume relationship of the material [19].

According to Figure 7, in the experiment conducted under ME conditions, the $a$ and $c$ lattice parameters of the α-phase (black circles) exhibit slightly greater compressibility than in the other two experiments (red and blue circles). This is correspondingly reflected in a marginally higher compressibility of the unit-cell volume in the ME experiment.

The pressure dependence of the unit-cell volume was analyzed using a third-order Birch–Murnaghan equation of state (EOS) [20]. In the fitting procedure, the ambient-pressure unit-cell volume ($V_0$) was fixed at the experimentally measured value of 17.39 $Å^3$. The resulting bulk modulus ($K_0$) and its pressure derivative ($K_0'$) are as follows: for the ME experiment, $K_0$ = 113(4) GPa and $K_0'$ = 4.04(6); for the SO experiment, $K_0$ = 120(5) GPa and $K_0'$ = 3.8(3); and for the MgO experiment, $K_0$ = 116(4) GPa and $K_0'$ = 4.1(3). Within experimental uncertainty, all three measurements yield $K_0'$ values consistent with 4. A slight increase in the bulk modulus is observed, from 113(4) GPa in the ME experiment to 120(5) GPa in the SO experiment. Comparison with titanium shows that the bulk modulus of Ti3Al2.5V agrees within error with that of pure Ti ($K_0$ = 119(9) GPa [8]). These values are also comparable to those reported for Ti64, $K_0$ = 101–125 GPa [10], although a higher value of 154 GPa has been reported for strongly non-hydrostatic conditions in experiments conducted without a pressure-transmitting medium. At the phase transition observed in Ti3-2.5 we determined that there is a small volume collapse (~1.7%) which is consistent with the first-order character reported for the α-ω transition in Ti and Ti64, reinforcing the idea that the transformation mechanism remains fundamentally similar across the alloy series.

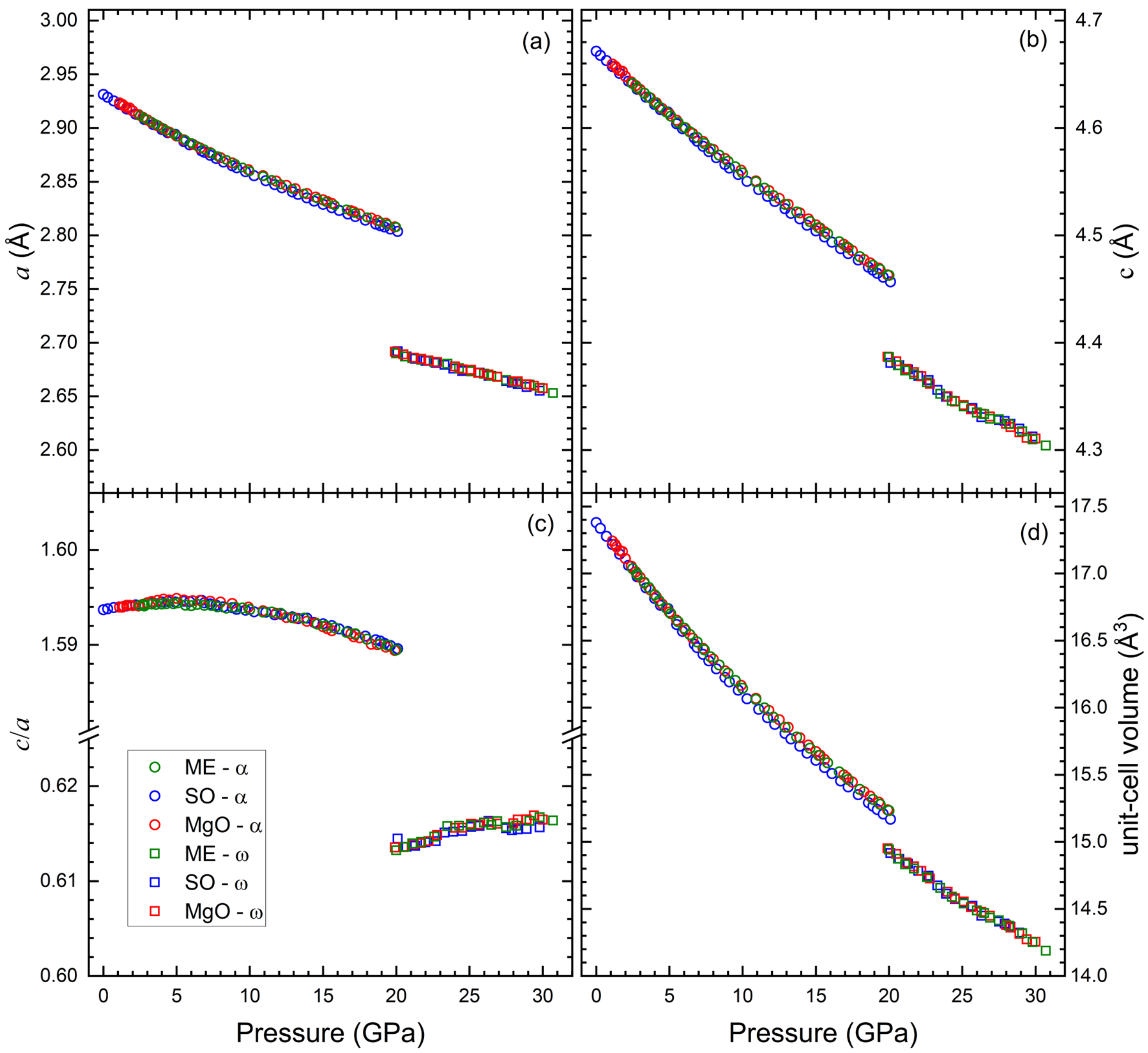

**Figure 7:** Pressure dependence of the unit-cell parameters of α-and ω-phase Ti3-2.5; (a) for *a* and (b) for *c*, the c/a ratio (c), and the unit-cell volume per formula unit (d). Circles show results for the α-phase and squares the results for the ω-phase. Green symbols are from the experiment performed on Ti3-2.5 embedded in an ME PTM, blue symbols for a SO PTM, and red symbols for an MgO PTM. Pressure media and phases are also identified in the inset of the figure.

Regarding the evolution of lattice parameters of the α-phase, we found a non-monotonic behavior of the c/a ratio (see Fig. 7(c)) which reaches a maximum value around 8 GPa and then gradually decreases with pressure. The behavior is qualitatively similar to that of Ti [8] and Ti64 [11]. In contrast, in the ω-phase there is a linear increase of c/a as also reported for Ti [8] and Ti64 [10, 11].

For the ω-phase, the volume data exhibit greater scatter than in the α-phase, preventing a clear distinction between the pressure–volume trends of the three experiments. Therefore, a single fit was performed using the combined dataset from all experiments. A third-order Birch–Murnaghan equation of state (EOS) was employed, with $V_0$, $K_0$, and $K_0'$ as free parameters. The fit yields $V_0$=16.9(1) Å$^3$, $K_0$=134(1) GPa, and $K_0'$=3.8(2). The zero-pressure volume is approximately 2.8% smaller than that of the α phase, consistent with previous results for Ti [8] and Ti64 [10,11]. The bulk modulus is about 12% higher than in the α phase. The obtained values of $K_0$ and $K_0'$ are also very close to those reported for Ti - $K_0$=138(10) GPa and $K_0'$=3.8(5). For Ti64, values of $K_0$=115(8) GPa and $K_0'$=4.6(5) [11] have been reported in a different pressure range. The differences between these parameters and the present results correspond to less than ~2.5σ for $K_0$ and ~1.5σ for $K_0'$, indicating that the two sets of results are statistically consistent within uncertainties.

Based on our results we conclude that, despite the strong dependence of transition pressure on composition, the equation-of-state results indicate that the bulk modulus remains nearly constant within experimental uncertainty across Ti, Ti3-2.5, and Ti64. This apparent decoupling between compressibility and phase stability is a key finding. It implies that while alloying significantly modifies the relative free energies of competing crystal structures (thus affecting transition pressures), it has a comparatively minor effect on the curvature of the energy–volume relationship near equilibrium (which governs the bulk modulus). The present results therefore suggest that alloying primarily modifies phase stability rather than elastic response, enabling the tuning of pressure-induced transformations without substantially affecting mechanical stiffness. This distinction provides a useful framework for understanding and designing Ti-based alloys for applications under extreme conditions. In other words, alloying alters phase stability more strongly than it alters elastic stiffness.

Taken together, these results establish that Ti3-2.5 behaves as an intermediate system bridging pure Ti and Ti64, both in terms of transition pressure and transformation behavior. The findings highlight that alloying provides an effective means to tune structural stability under extreme conditions without significantly altering elastic properties, which is particularly relevant for applications where both mechanical integrity and phase stability are critical.

## Conclusion

High-pressure XRD experiments on Ti3-2.5 reveal an $\alpha$-$\omega$ phase transition between 17–19 GPa, placing this alloy between pure Ti and Ti64. Our results provide direct experimental confirmation that the transition pressure increases with combined Al and V content in Ti-Al-V alloys. While pressure-transmitting media affect the exact transition pressure and reversibility, the overall compositional trend remains robust. Equation-of-state results show that the bulk modulus is nearly unchanged across Ti–Al–V alloys, indicating a decoupling between compressibility and phase stability. Thus, alloying primarily tunes transition pressure without significantly affecting elastic stiffness. Overall, Ti3-2.5 behaves as an intermediate system, demonstrating that composition provides an effective means to control high-pressure phase stability in Ti–Al–V alloys with implications for the design of titanium alloys intended to operate under extreme mechanical conditions. Based on our results, the high-pressure behavior of other Ti–Al–V alloys can be predicted; for example, commercial Ti-5Al-3V is expected to undergo an $\alpha \rightarrow \omega$ transition at approximately 22–25 GPa, following the systematic compositional trend in Ti–Al–V alloys.

## Disclosure statement

No potential conflict of interest was reported by the authors.

## Data availability statement

The data that support the findings of this study are available from the corresponding author upon reasonable request.

## Additional information

### Funding

This work was carried out with the financial support from the Spanish Research Agency (AEI) and Spanish Ministry of Science, Innovation, and Universities (MCIU) under grant PID2022-138076NB-C41 (DOI: 10.13039/501100011033). P.B. and D.E. thank GVA for the Postdoctoral Fellowship No. CIAPOS/2023/406.

## Acknowledgments

The XRD experiments were performed at the MSPD-BL04 beamline at ALBA Synchrotron with the collaboration of ALBA staff, under proposal number 20250370105.

## References

[1] Chen X, Xie Y, Zhang T, et al. Harnessing strengthening-metastability synergy for extreme work hardening in additively manufactured titanium alloys, Nat. Commun. 2026, 17, 959 . DOI: 10.1038/s41467-025-67683-8

[2] Ishida T, Kano S, Wakai E, *et al.* Contrasting irradiation behavior of dual phases in Ti-6Al-4V alloy at low-temperature due to ω-phase precursors in β-phase matrix. J. Alloys Compd. 2024, 995, 174701. DOI: 10.1016/j.jallcom.2024.174701

[3] Leguey T, Schäublin R, Marmy P, and Victoria M, Microstructure of Ti5Al2.5Sn and Ti6Al4V deformed in tensile and fatigue tests. J. Nuclear Mat. 2002, 305, 52-59. DOI: 10.1016/S0022-3115(02)00888-7

[4] Bartolomeu F, Gasik M, Silva FS, and Miranda G, Mechanical Properties of Ti6Al4V Fabricated by Laser Powder Bed Fusion: A Review Focused on the Processing and Microstructural Parameters Influence on the Final Properties. Metals 2022, 12, 986. DOI: 10.3390/met12060986

[5] Parakh A, Lee AC, Chariton S, *et al.* High pressure deformation induced precipitation in Al–Zn–Mg–Cu alloy (Al7075). Materials Science and Engineering A 2022, 853, 143765. DOI: 10.1016/j.msea.2022.143765

[6] Errandonea D, Burakovsky L, Preston DL, *et al.* Experimental and theoretical confirmation of an orthorhombic phase transition in niobium at high pressure and temperature. Commun Mater 2020, 1, 60. DOI: 10.1038/s43246-020-00058-2

[7] Smit, D, Joris OPJ, Sankaran A, *et al.* On the high-pressure phase stability and elastic properties of β-titanium alloys. J. Phys.: Condens. Matter 2017, 29, 155401. DOI: 10.1088/1361-648X/aa60b6

[8] Errandonea D, Meng Y, Somayazulu M, Häusermann D, Pressure-induced $\alpha$-$\omega$ transition in titanium metal: a systematic study of the effects of uniaxial stress. Physica B 2005, 355, 116-125. DOI: 10.1016/j.physb.2004.10.030

[9] Dewaele A, Stutzmann V, Bouchet J, *et al.* High pressure-temperature phase diagram and equation of state of titanium. Phys. Rev. B 2015, 91, 134108. DOI: 10.1103/PhysRevB.91.134108

[10] MacLeod S, Tegner BE, Cynn H, *et al.* Experimental and theoretical study of Ti-6Al-4V to 220 GPa. Phys. Rev B 2012. 85, 224202. DOI: 10.1103/PhysRevB.85.224202

[11] MacLeod SG, Errandonea D, Cox GA, *et al.* The phase diagram of Ti-6Al-4V at high-pressures and high-temperatures. J. Phys.: Condens. Matter 2021, 33, 154001. DOI: 10.1088/1361-648X/abdffa

[12] Klotz S, Chervin JC, Munsch P, and Le Marchand G, Hydrostatic limits of 11 pressure transmitting media. J. Phys. D: Appl. Phys. 2009, 42, 075413. DOI 10.1088/0022-3727/42/7/075413

[13] Speziale S, Zha CS, Duffy TS, *et al.* Quasi-hydrostatic compression of magnesium oxide to 52 GPa: Implications for the pressure-volume-temperature equation of state. J. Geophysical Research 2001, 106(B1), 515–528. DOI: 10.1029/2000JB900318

[14] Fauth F, Peral I, Popescu C, and Knapp M, The new Material Science Powder Diffraction beamline at ALBA Synchrotron. Powder Diffr. 2013, 28, S360–S370. DOI: 10.1017/S0885715613000900

[15] Dewaele A, Loubeyre P, and Mezouar M, Equations of state of six metals above 94GPa. Phys. Rev. B 2004, 70, 094112. DOI: 10.1103/PhysRevB.70.094112

[16] Le Bail A, Duroy H, and Fourquet JL, Ab-initio structure determination of $LiSbWO_6$ by X-ray powder diffraction. Mat. Res. Bull. 1988, 23, 447-452. DOI: 10.1016/0025-5408(88)90019-0

[17] Hennig RG, Trinkle DR, Bouchet J, *et al.* mpurities block the $\alpha$ to $\omega$ martensitic transformation in titanium. Nature Materials 2005, 4, 129–133. DOI: 10.1038/nmat1292

[18] Williams JC, and Starke Jr. EA, Progress in structural materials for aerospace systems. Acta Materialia 2003, 51, 5775–5799. DOI: 10.1016/j.actamat.2003.08.023

[19] Konôpková Z, Rothkirch A, Singh AK, *et al.*, In situ x-ray diffraction of fast compressed iron: Analysis of strains and stress under non-hydrostatic pressure. Phys. Rev. B 2015, 91, 144101. DOI: 10.1103/PhysRevB.91.144101

[20] Birch F, Finite Elastic Strain of Cubic Crystals. Phys. Rev. 1947, 71, 809–824. DOI: 10.1103/PhysRev.71.809